\documentclass[aps,floats,showpacs,preprint]{revtex4} 
\usepackage{graphicx}
\usepackage{epsfig}
\usepackage{psfig}
\usepackage{amsmath}
\usepackage{hyperref}
\def\PRL#1{{ Phys.\ Rev.\ Lett.} {\bf #1}}
\def\PRD#1{{ Phys.\ Rev.} {\bf D#1}}
\def\NPB#1{{ Nucl.\ Phys.} {\bf B#1}}
\def\PLB#1{{Phys.\ Lett.} {\bf B#1}}

\def\be{\begin{equation}}
\def\ee{\end{equation}}
\def\bea{\begin{eqnarray}}
\def\eea{\end{eqnarray}}

{\newcommand{\lsim}{\mbox{\raisebox{-.6ex}{~$\stackrel{<}{\sim}$~}}}
{\newcommand{\gsim}{\mbox{\raisebox{-.6ex}{~$\stackrel{>}{\sim}$~}}}
\def\sss{\scriptscriptstyle}

\def\GD{\Gamma_{\sss D}}
\def\GS{\Gamma_{\sss S}}
\def\TBL{T_{B-L}}
\def\non{N_1}
\def\Ls{{\sss L}}

\begin{document}

\title{Gauged $B-L$ symmetry and baryogenesis via leptogenesis at TeV scale}

\author{Narendra Sahu}
\email{narendra@phy.iitb.ac.in}
\author{Urjit A. Yajnik}
\email{yajnik@phy.iitb.ac.in}
\affiliation{Department of Physics, Indian Institute of Technology,
Bombay, Mumbai 400076 INDIA}                   

\begin{abstract}
It is shown that the requirement of preservation of baryon asymmetry
does not rule out a scale for leptogenesis as low as $10$ TeV.  The 
conclusions are compatible with see-saw mechanism if for example 
the pivot mass scale for neutrinos is  $\approx\, 10^{-2}$ that of the 
charged leptons. We explore the parameter space $\tilde{m}_1$-
$M_1$ of relevant light and heavy neutrino masses by solving
Boltzmann equations. A viable scenario for 
obtaining baryogenesis in this way is presented in the
context of gauged $B-L$ symmetry.
\end{abstract}
\maketitle

\section{Introduction}
It has long been recognized that existence of heavy Majorana 
neutrinos has important consequences for the baryon asymmetry 
of the Universe. With the discovery of the neutrino masses and 
mixings, it becomes clear that only $B-L$ can be considered to 
be a conserved global symmetry of the Standard Model (SM) and 
not the individual quantum numbers $B-L_e$, $B-L_\mu$ and 
$B-L_\tau$. Combined with the fact that the classical $B+L$ 
symmetry is anomalous~\cite{krs.86, arn_mac.88,aaps.91} it 
becomes important to analyse the consequences of any $B-L$ 
violating interactions because the two effects combined can 
result in the unwelcome conclusion of the net baryon 
number of the Universe being zero.

At present two broad possibilities exist as viable explanations 
of the observed baryon asymmetry of the Universe. One is the 
baryogenesis through leptogenesis~\cite{fukugita.86}. This has 
been analysed extensively in~\cite{luty.92, mohapatra_prd.92,
plumacher.96,buch_bari_plum.02} and has provided very robust 
conclusions for 
neutrino physics. Its typical scale of operation has to be high, 
at least an intermediate scale of $10^9$ GeV. This has to do with 
the intrinsic competition needed between the lepton number 
violating processes and the expansion scale of the Universe. 
While the mechanism does not prefer any particular unification 
scheme, it has the virtue of working practically unchanged
upon inclusion of supersymmetry~\cite{plumacher.97}. 

The alternative to this is provided by mechanisms which  work 
at the TeV scale~\cite{bou.02,bou_ham_sen_prl.04,pil_und.04,
ham_rus_west.04} and rely on the new particle content implied in 
supersymmetric extensions of the SM. The Minimal Supersymmetric
SM (MSSM) holds only  marginal possibilities for baryogenesis. 
The Next to Minimal or NMSSM possesses robust mechanism for 
baryogenesis~\cite{huber&schimdt.01} however the model has 
unresolved issues vis a vis the $\mu$ problem due to domain 
walls~\cite{abel_sar_white.95}. However its restricted version, 
the nMSSM is reported~\cite{pan_tam_PLB1.99,pan_tam_PLB2.99,
pan&pil_PRD.01,ded_hug_mor_tam_PRD.01,men_mor_wag_prd.04} 
to tackle all of the concerned issues.

It is worth investigating other possibilities, whether or not 
supersymmetry is essential to the mechanism. In this paper we 
study afresh the consequence of heavy Majorana neutrinos given 
the current knowledge of light neutrinos. The starting point 
is the observation~\cite{har_tur.90,fglp.91} that the 
heavy neutrinos participate in the erasure of any pre-existing 
asymmetry through scattering as well as decay  and inverse decay 
processes. Estimates using general behavior of the thermal 
rates lead to a conclusion that there is an upper bound on the 
temperature $T_{\rm B-L}$ at which $B-L$ asymmetry could have
been created. This bound is 
$T_{B-L}\lsim 10^{13}$GeV$\times(1 eV/m_\nu)^2$,
where $m_\nu$ is the typical light neutrino mass. This bound is
too weak to be of accelerator physics interest. We extend this 
analysis by numerical solution of the Boltzmann equations and 
obtain regions of viability in the parameter space spanned by 
$\tilde{m}_1$-$M_1$, where $\tilde{m}_1$ is a light neutrino mass 
parameter defined in eq. (\ref{eq:mtilde}) and $M_1$
is the mass of the lightest of the heavy Majorana neutrinos.  
We find that our results are in consonance with~\cite{fglp.91} 
where it was argued that scattering processes provide a weaker 
constraint than the decay processes. If the scatterings
become the main source of erasure of the primordial asymmetry 
then the constraint can be interpreted to imply  $\TBL<M_1$. 
Further, this temperature can be as low as TeV range with 
$\tilde{m}_1$ within the range expected from neutrino 
observations. This is compatible with  see-saw mechanism if
the "pivot" mass  scale is that of the charged leptons. 

In~\cite{tyatgat_plb.93,cynr.02} it was shown that the 
Left-Right symmetric 
model~\cite{moh&sen.81,moh_susy_book.92} presents just
such a possibility. In this model $B-L$ appears as a gauged 
symmetry in a natural way. The phase transition is rendered 
first order so long as there is an approximate discrete symmetry 
$L\leftrightarrow R$, independent of details of other parameters. 
Spontaneously generated CP phases then allow creation of lepton 
asymmetry.  We check this scenario here against our numerical 
results and in the light of the discussion above.

In the following we first recapitulate the arguments concerning
erasure of $B-L$ asymmetry due to heavy Majorana neutrinos, 
then present the relevant numerical results, then discuss the 
possibility for the Left-Right symmetric model and present a 
summary in conclusion.

\section{Erasure constraints - simple version}
The presence of several heavy Majorana neutrino species ($N_i$) 
gives rise to processes depleting the existing lepton 
asymmetry in two ways. They are (i) scattering processes (S) 
among the SM fermions and (ii) Decay (D) and inverse decays (ID) 
of the heavy neutrinos. We assume a normal hierarchy among the 
right handed neutrinos such that only the lightest of the right 
handed  neutrinos ($\non$) makes a significant contribution to
the above mentioned processes. 
At first we use a simpler picture, though 
the numerical results to follow are based on the complete 
formalism. The dominant contributions to the two types of 
processes are governed by the temperature dependent rates
\be
\label{rates}
\GD \sim {h^2 M_1^2 \over 16\pi(4T^2 + M_1^2 )^{1/2}}
\hspace{1cm} {\rm and} \hspace{1cm}
\GS \sim {h^4 \over 13\pi^3}{T^3 \over 
(9T^2 + M_1^2)},
\ee
where $h$ is typical Dirac Yukawa coupling of the neutrino.

Let us first consider the case  $M_1>\TBL$. For $T<\TBL$, the $\non$ 
states are not populated, nor is there sufficient free energy to 
create them, rendering the D-ID processes unimportant. 
We work in the scenario where the sphalerons are in equilibrium,
maintaining rough equality of $B$ and $L$ numbers. As the $B-L$
continues to be diluted we estimate the net baryon asymmetry
produced as~\cite{cynr.02}
\be
10^{-d_{\sss B}}\equiv\exp\left(-\int_{t_{B-L}}^{t_{\sss EW}} \GS dt\right)
=\exp\left(-\int_{T_{\sss EW}}^{\TBL} {\GS\over H}\,{dT\over T}\right),
\ee
where $t_{B-L}$ is the time of the $(B-L)$-breaking phase transition, 
$H$ is the Hubble parameter, and $t_{\sss EW}$  and $T_{\sss EW}$ 
corresponds to the electroweak scale after which the sphalerons 
are ineffective. Evaluating the integral gives an estimate for 
the exponent as 
\be
\label{dsB}
	d_{\sss B} \cong {3\sqrt{10}\over 13\pi^4\ln10\sqrt{g_*}}\, 
h^4{M_{Pl} \TBL\,\over M_1^2}.
\ee
The same result upto a numerical factor is obtained 
in~\cite{buch_bari_plum.03}, the suppression factor $\omega^{(2)}$ 
therein. Eq.\ (\ref{dsB}) can be solved for the Yukawa coupling 
$h$ which gives the Dirac mass term for the neutrino: 
$h^4 \lsim 3200\, d_{\sss B}\left({M_1^2\over \TBL M_{Pl}}
\right)$ where we have taken $g_* = 110$ for definiteness and
$d_{\sss B}$ here stands for total depletion including 
from subdominant channels. 
This can be further transformed into an upper limit on the 
light neutrino masses using the canonical seesaw relation.
Including the effect of generation mixing,
the parameter  which appears in the thermal rates is 
\be
\label{eq:mtilde}
\tilde{m_1} \equiv \frac{(m_D^{\dagger} m_D)_{11}}{M_1}
\ee
and is called the \emph{effective neutrino mass}~\cite{plumacher.96}.
The constraint (\ref{dsB}) can then be recast as
\be
\label{eq:mnudb}
m_\nu\ \sim\  \tilde{m_1}  \lsim {180 v^2\over \sqrt{\TBL  M_{Pl}}}\,\left({d_{\sss
B}\over 10}\right)^{1/2}
\ee
This bound is useful for the case of large suppression. Consider
 $d_{\sss B}=10$.  If we seek $\TBL\sim 1$TeV and $M_1\sim10$TeV, 
 this bound is saturated for $m_\nu\approx 50$keV,
corresponding to $h\approx m_\tau/v$. This bound is academic in view of
the WMAP bound $\sum m_{\nu_i}\approx 0.69 eV$ 
\cite{ParamWMAP} . On the other hand, 
for the phenomenologically  interesting  case $m_\nu \approx 10^{-2}$eV, 
with $h\approx 10^{-5}\approx m_e/v$ and with $M_1$ and $\TBL$ as above,
eq. (\ref{eq:mnudb})      can be read to imply that in fact $d_{\sss B}$ 
is vanishingly small.
This in turn demands, in view of the low scale we are seeking, a 
non-thermal mechanism for producing lepton asymmetry naturally in the 
range $10^{-10}$. Such a mechanism is discussed in sec. \ref{sec:lasymLR}.

In the opposite regime $M_1<\TBL$, both of the above types of 
processes could freely occur. The condition that complete erasure 
is prevented requires that the above processes are slower than 
the expansion scale of the Universe for all $T>M_1$. It turns out 
to be sufficient~\cite{fglp.91} to require $\GD<H$ which also 
ensures that $\GS<H$. This leads to the requirement 
\be
m_\nu < m_* \equiv 4\pi g_*^{1/2}\frac{G_N^{1/2}}{\sqrt{2}G_F}
= 6.5\times10^{-4}eV
\label{FGLPbound}
\ee
where the parameter $m_*$~\cite{fglp.91} contains only universal
couplings and $g_*$, and may be called the \emph{cosmological neutrino 
mass}. 

The constraint of equation (\ref{FGLPbound}) is compatible 
with models of neutrino mass if we identify the neutrino 
Dirac mass matrix $m_D$ as that of charged leptons. For a 
texture of $m_D$~\cite{fritzsch.79,sahu&uma_prd.04} 
\be
m_D=\frac{m_{\tau}}{1.054618}
\begin{pmatrix}
0 & a & 0\\
a & 0 & b\\
0 & b & c\end{pmatrix},
\label{texture}
\ee 
the hierarchy for charged leptons can be generated if 
\begin{equation}
a=0.004,~~~ b=0.24 ~~~{\mathrm and}~~~ c=1.
\label{abcvalues}
\end{equation}
For these set of values of a, b, and c the above mass matrix 
$m_D$ is normalized with respect to the $\tau$-lepton mass. 
Using the mass matrix (\ref{texture}) in equation 
(\ref{eq:mtilde}) we get   
\be
\tilde{m_1}=4.16\times 10^{-4}eV \left(\frac{10^8 GeV}{M_1}\right).
\ee
Thus with this texture of masses, eq. (\ref{FGLPbound}) is satisfied
for $M_1\gsim 10^8 GeV$. If we seek $M_1$ mass within the TeV range,
this formula suggests that the texture for the neutrinos should have
the Dirac mass scale smaller by $10^{-2}$ relative to the charged
leptons.

The bound (\ref{FGLPbound}) is meant to ensure that depletion effects
remain unimportant and is rather strong. A more detailed estimate 
of the permitted values of $\tilde{m}_1$ and $M_1$ is obtained by 
solving the relevant Boltzmann equations. In the rest of the paper 
we investigate these for the case $\TBL>M_1$ and $\Gamma_D<H$. 

\section{Solution of Boltzmann equations}
The relevant Boltzmann equations for our purpose are
~\cite{luty.92,plumacher.96, buch_bari_plum.02}
\bea
\frac{dY_{N1}}{dZ} &=& -(D+S)\left(Y_{N1}-Y^{eq}_{N1}\right)
\label{boltzmann.1}\\
\frac{dY_{B-L}}{dZ} &=& -W Y_{B-L}
\label{boltzmann.2},
\eea
where $Y_i=(n_{i}/s)$ is the density of the species 
$i$  in a comoving volume, $Z=M_1/T$ and $s=43.86(g_*/100)T^3$ is 
the entropy density at an epoch of temperature T. 
The two terms D and S on the right hand side of 
equation (\ref{boltzmann.1}) change the density of $N_1$ 
in a comoving volume while the right hand side of equation 
(\ref{boltzmann.2}) accounts for the wash out or dilution effects. 
We assume no new processes below $T_{B-L}$ which can 
create lepton asymmetry. In a recent work~\cite{strumia.04} it has 
been reported that the thermal corrections to the above processes 
as well as the processes involving the gauge bosons are important 
for final L-asymmetry. However their importance is under 
debate~\cite{buch-bari-plum-ped.04}. In this paper we limit 
ourselves to the same formalism as in~\cite{plumacher.96,
buch_bari_plum.02}. In equation (\ref{boltzmann.1}) 
$D=\Gamma_D/HZ$, where $\Gamma_D$ determines the decay rate 
of $N_1$, $S=\Gamma_S/HZ$, where $\Gamma_S$ determines the rate 
of $\Delta_{\rm L}=1$ lepton violating scatterings. In equation 
(\ref{boltzmann.2}) $W=\Gamma_W/HZ$, where $\Gamma_W$ incorporates 
the rate of depletion of the B-L number involving the lepton 
violating processes with $\Delta_{\rm L}=1$, $\Delta_{\rm L}=2$ 
as well as inverse decays creating $N_1$. The various $\Gamma$'s 
are related to the scattering densities~\cite{luty.92} $\gamma$s as 
\begin{equation}
\Gamma_i^X(Z)=\frac{\gamma_i(Z)}{n_X^{eq}}
\end{equation}
The dependence of the scattering rates involved in
$\Delta_{\rm L}=1$ lepton violating processes on the 
parameters $\tilde{m}_1$ and $M_1$ is similar  to that of the 
decay rate $\Gamma_{D}$. As the Universe expands these 
$\Gamma$'s compete with the Hubble expansion parameter. 
Therefore in a comoving volume we have 
\be
\left(\frac{\gamma_{D}}{sH(M_1)}\right), \left(\frac{
\gamma^{N1}_{\phi,s}}{sH(M_1)}\right), \left(\frac{
\gamma^{N1}_{\phi,t}}{sH(M_1)}\right) \propto k_1\tilde{m}_1.
\label{dilution}
\ee
On the other hand the dependence of the $\gamma$'s in 
$\Delta_{\rm L}=2$ lepton number violating processes on 
$\tilde{m}_1$ and $M_1$ are given by 
\be
\left(\frac{\gamma^l_{N1}}{sH(M_1)}\right), \left(\frac{
\gamma^l_{N1,t}}{sH(M_1)}\right) \propto k_2 \tilde{m}_1^2 M_1.
\label{washout}
\ee
Finally there are also lepton conserving processes 
where the dependence is given by 
\be
\left(\frac{\gamma_{Z'}}{sH(M_1)}\right) \propto k_3 M_1^{-1}.
\label{l-conserve}
\ee
In the above equations (\ref{dilution}), (\ref{washout}), (\ref{l-conserve}),  
$k_i$, $i=1,2,3$ are dimensionful constants determined from other
parameters. Since the 
lepton conserving processes are inversely proportional 
to the mass scale of $N_1$, they rapidly bring the species $N_1$ into 
thermal equilibrium for $M_1<T$.
Further, for  the smaller values of $M_1$, the washout effects 
(\ref{washout}) are negligible because of their linear 
dependence on $M_1$. This is the regime in which we are
while solving the Boltzmann equations in the following.  

\begin{figure}[ht]
\begin{center}
\epsfig{file=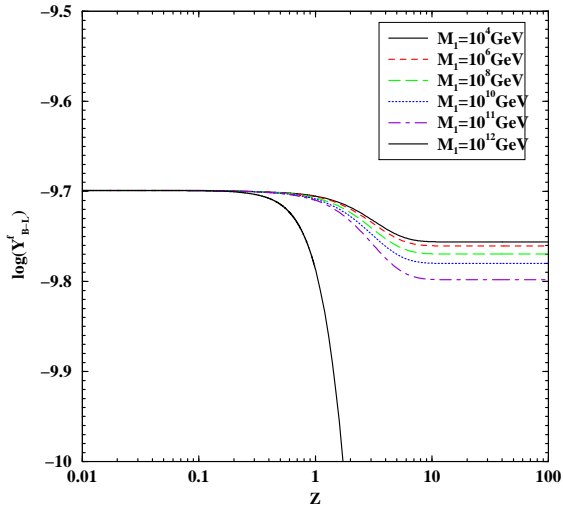, width=0.45\textwidth}
\caption{The evolution of B-L asymmetry for different values of 
$M_1$ shown against $Z(=M_1/T)$ for $\tilde{m}_1=10^{-4}$eV 
and $\eta^{raw}=2.0\times 10^{-10}$}
\label{figure-1}
\end{center}
\end{figure}

\begin{figure}[ht]
\begin{center}
\epsfig{file=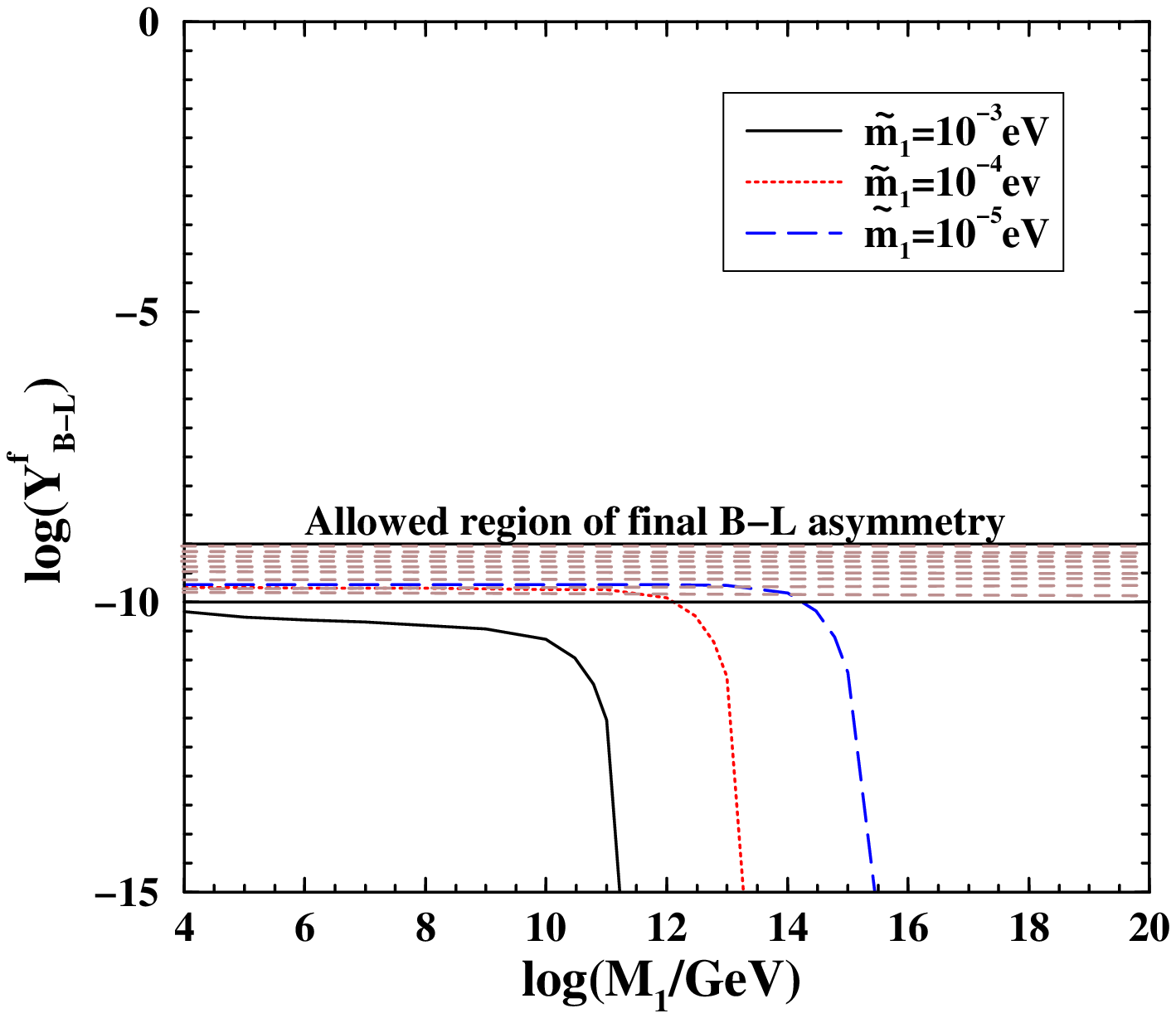, width=0.45\textwidth}
\caption{The allowed values of $M_1$ against the
required final asymmetry is shown for $\eta^{raw}=
2.0\times 10^{-10}$}
\label{figure-2}
\end{center}
\end{figure}  
\begin{figure}[hbt]
\begin{center}  
\epsfig{file=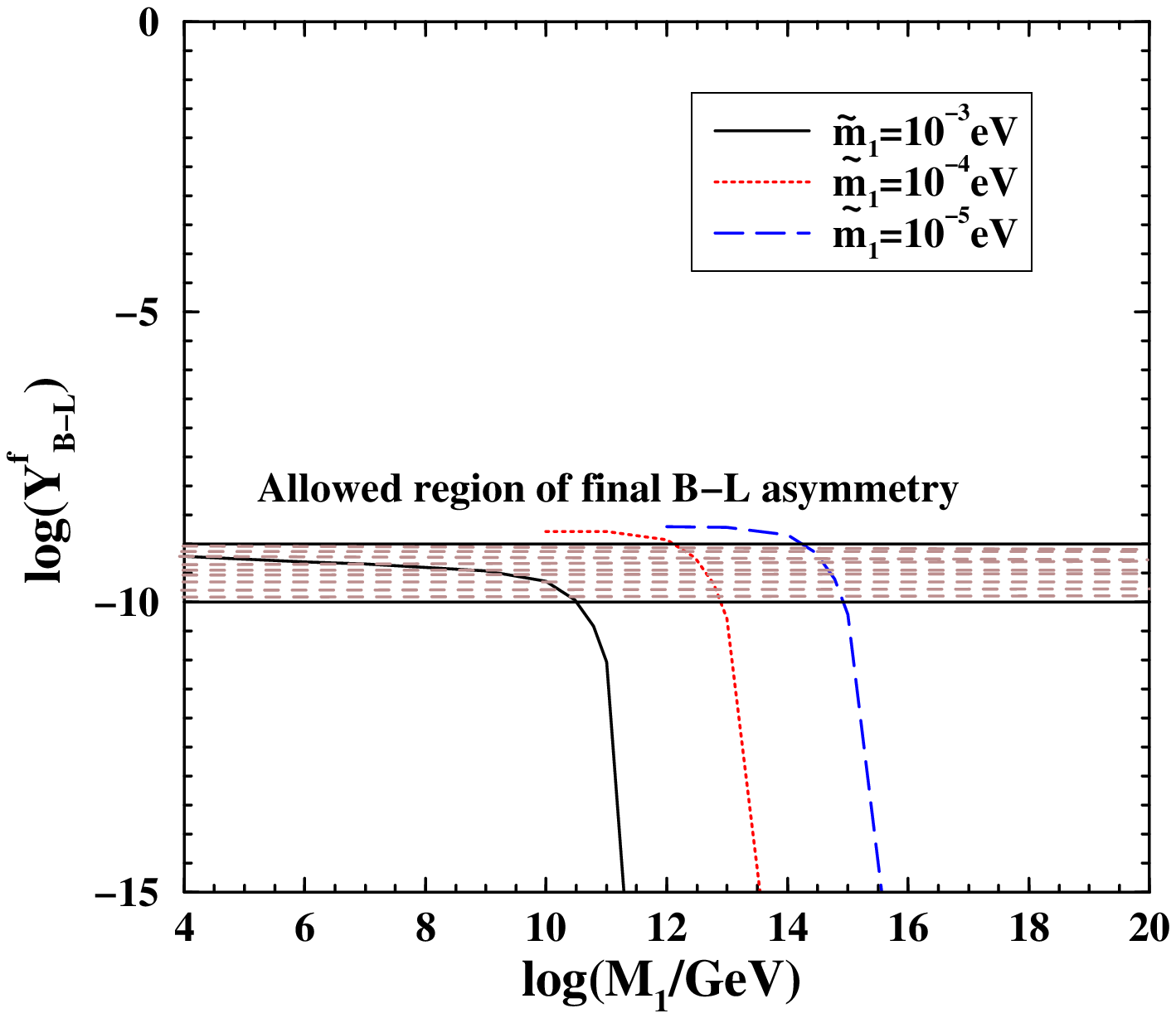, width=0.45\textwidth}
\caption{The allowed values of $M_1$ against the final 
required asymmetry is shown for $\eta^{raw}=2.0\times 10^{-9}$}
\label{figure-3}
\end{center}
\end{figure}

\begin{figure}[hbt]
\begin{center}
\psfig{file=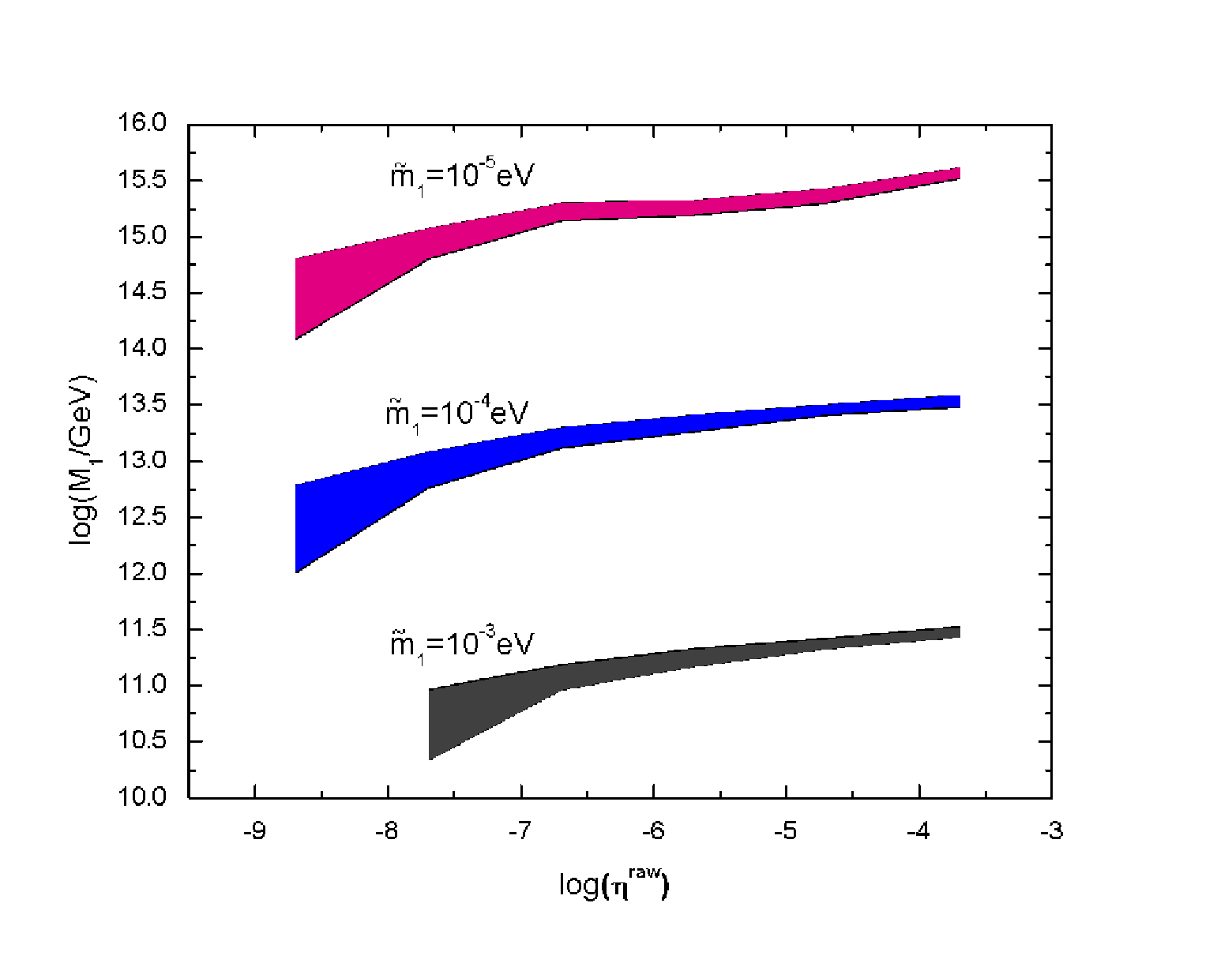, width=0.5\textwidth}
\caption{The allowed region of $M_1$ is shown for different 
values of $\tilde{m}_1$ for large values of $\eta^{\sss raw}$}
\label{fig:regions}
\end{center}
\end{figure}   

The equations (\ref{boltzmann.1}) and (\ref{boltzmann.2}) are solved
numerically. The initial $B-L$ asymmetry is the net raw 
asymmetry produced during the B-L symmetry breaking phase 
transition by any thermal or non-thermal process. As such we 
impose the following initial conditions  
\be
Y^{in}_{N1}=Y^{eq}_{N1}~~ {\mathrm and}~~ Y^{in}_{B-L}=\eta^{raw}_{B-L}.
\label{initial-cond}
\ee 

At temperature $T\geq M_1$, wash out effects involving $N_1$ are 
kept under check due to the $\tilde{m}_1^2$ dependence in (\ref{washout}) 
for small values of $\tilde{m}_1$. As a result a given 
raw asymmetry suffers limited erasure. As the temperature 
falls below the mass scale of $M_1$ the wash out processes 
become negligible leaving behind a final lepton asymmetry. 
Fig. \ref{figure-1} shows the result of solving the Boltzmann equations 
for different values of $M_1$.  

If we demand that the initial raw asymmetry is of the order 
of $\alpha \times 10^{-10}$, with $\alpha\sim O(1)$, then 
in order to preserve the final asymmetry of the same order 
as the initial one it is necessary that the neutrino mass 
parameter $\tilde{m}_1$ should be less than $10^{-3}eV$. 
This can be seen from fig. \ref{figure-2}. For $\tilde{m}_1= 10^{-3}eV$ we 
can not find any value of $M_1$ to preserve the final asymmetry, 
$\alpha \times 10^{-10}$ in the allowed region. This is because 
of the large wash out effects as inferred from the equation 
(\ref{washout}). However, for $\tilde{m_1}= 10^{-4}eV$ we get 
a {\it lowest threshold} on the lightest right handed neutrino 
of the order $10^{12}GeV$. For any value of $M_1\leq 10^{12}GeV$ the 
final asymmetry lies in the allowed region. This bound increases by 
two order of magnitude for further one order suppression of 
the neutrino parameter $\tilde{m}_1$. The important point 
being that $M_1=10 TeV$ is within the acceptable range.
                                                        
We now consider the raw asymmetry one order more than the 
previous case {\it i.e.} $\eta^{\sss raw}_{\sss B-L}= \alpha 
\times 10^{-9}$. From fig. \ref{figure-3} we see that 
for $\tilde{m}_1=10^{-3}eV$ there is only an upper bound 
$M_1= 10^{10.5}GeV$, such that the final asymmetry lies in the 
allowed region for all smaller values of $M_1$. Thus for the case 
of raw asymmetry an order of magnitude smaller, the upper bound 
on $M_1$ decrease by two orders of magnitude (e.g. compare previous 
paragraph). However, the choice of smaller values of $\tilde{m}_1$ 
leads to a small window for values of $M_1$ for which we end up 
with the final required asymmetry. In particular for 
$\tilde{m}_1= 10^{-4}eV$ the allowed range for $M_1$ 
is ($10^{12}~ - ~10^{13})GeV$, while for $\tilde{m}_1= 
10^{-5}eV$ the allowed range shifts to ($10^{14}~ - 
~10^{15})GeV$. The window effect can be understood as 
follows. Increasing the value of $M_1$ tends to lift the 
suppression imposed by the $\tilde{m}_1^2$ dependence 
of the wash out effects, thus improving efficiency of 
the latter. However, further increase in $M_1$ makes the 
effects too efficient, erasing the raw asymmetry to 
insignificant levels.

The windowing effect emerges clearly as we consider the cases 
of large raw asymmetries. This is shown in fig. \ref{fig:regions}. 
It is seen that as the raw asymmetry increases the allowed regions 
become progressively narrower and lie in the range $(10^{10}~-~10^{15})
GeV$. Thus a given raw lepton asymmetry determines a corresponding 
small range of the heavy Majorana neutrino masses for which we can 
obtain the final asymmetry of the required order $\alpha\times 
10^{-10}$. Again smaller is the effective neutrino mass $\tilde{m}_1$ 
larger is the mean value of the allowed mass of the heavy Majorana 
neutrino and this is a consequence of normal see-saw.      

Finally, in the following, we give an example for 
non-thermal creation of L-asymmetry in the context of 
left-right symmetric model. 

\section{Lepton asymmetry in left-right symmetric model}
\label{sec:lasymLR}
We discuss qualitatively the possibility of lepton 
asymmetry during the left-right symmetry breaking phase 
transition~\cite{cynr.02}. In the following we recapitulate 
the important aspects of left-right symmetric model 
for our purpose and the possible non-thermal mechanism of 
producing raw lepton asymmetry. This asymmetry which gets
converted to baryon asymmetry, can be naturally small if
the quartic couplings of the theory are small. Smallness
of zero-temperature $CP$ phase is not essential for this
mechanism to provide small raw $L$ asymmetry.

\subsection{Left-Right symmetric model and transient 
domain walls}
In the left-right symmetric model the right handed chiral leptons, 
which are singlet under the Standard Model gauge group 
$SU(2)_L\otimes U(1)_Y$, get a new member $\nu_R$ per 
family under the gauge group 
$SU(2)_L\otimes SU(2)_R\otimes U(1)_{B-L}$. The augmented symmetry 
in the Higgs sector requires two additional triplets $\Delta_L$ and 
$\Delta_R$ and a bidoublet $\Phi$, which contains two copies of $SM$ 
Higgs forming an $SU(2)_R$ representation. 

The Higgs potential of the theory naturally entails a vacuum
structure wherein at the first stage of symmetry breaking, either  
one of $\Delta_L$ or $\Delta_R$ acquires a vacuum expectation value 
the left-right symmetry, $SU(2)_L\leftrightarrow SU(2)_R$, breaks. 
It is required that $\Delta_R$ acquires a VEV first, resulting in
 $SU(2)_R \otimes U(1)_{B-L}$ $\rightarrow U(1)_Y$. Finally
$Q=T^3_L + T^3_R+ {1\over 2}(B-L)$, survives after  the bidoublet 
$\phi$ and the $\Delta_L$ acquire VEVs.                                                              

If the left-right symmetry were exact, the first stage of breaking 
gives rise to stable domain walls \cite{ywmmc,lazar,lew-rio} 
interpolating between the  L and R-like regions. 
By  L-like we mean regions favored by the observed phenomenology, 
while in the R-like regions the vacuum expectation value of 
$\Delta_R$ is zero. Unless some non-trivial mechanism prevents this
domain structure, the existence of R-like domains would disagree 
with low energy phenomenology. Furthermore, the domain walls would
quickly come to dominate the energy 
density of the Universe. Thus in this
theory a departure from exact symmetry in  such a way as to eliminate 
the R-like regions is essential. 

The domain walls formed can be transient if there exists a 
slight deviation from exact discrete symmetry. As a result the 
thermal perturbative corrections to the Higgs field free energy 
will not be symmetric and the domain walls will be unstable. 
This is possible if the low energy ($\sim10^4$GeV-$10^9$GeV)
left-right symmetric theory 
is descended from a Grand Unified Theory (GUT) and the effect is 
small, suppressed by the GUT mass scale.  In the process of cooling
the Universe would first pass through the phase transition
where this approximate symmetry breaks. The slight difference in
free energy between the two types of regions
causes a pressure difference across the walls,  converting all the R-like 
regions to L-like regions. Details of this dynamics can be found 
in ref.~\cite{cynr.02}.

\subsection{Leptogenesis mechanism}
At least two of the Higgs expectation values in L-R model
are generically complex, thus making it possible to achieve 
natural i.e., purely spontaneous $CP$
violation~\cite{dgko} permitting all parameters in the Higgs 
potential to be real. It was shown in \cite{cynr.02} that within 
the thickness  of the domain 
walls the net $CP$ violating phase becomes position
dependent. Under these circumstances a formalism exists~\cite{
jpt,clijokai,clikai}, wherein the chemical potential
$\mu_\Ls$ created for the Lepton number can be computed as 
a solution of the diffusion equation
\be
\label{eq:diffeq}
-D_\nu \mu_\Ls'' - v_w \mu_\Ls'
+ \theta(x)\, \Gamma_{\rm hf}\,\mu_\Ls = S(x).
\ee
Here $D_\nu$ is the neutrino diffusion coefficient,
$v_w$ is the velocity of the wall, taken to be moving in the $+x$
direction, $\Gamma_{\rm hf}$ is the rate of helicity
flipping interactions taking place in front of the wall (hence
the step function $\theta(x)$), and $S$ is the source term
which contains derivatives of the position dependent complex 
Dirac mass.
                                                      
After integration of the above equation and using inputs
from the numerical solutions we find the raw lepton
asymmetry~\cite{cynr.02}
\be
\eta^{\rm raw}_{\sss L} \cong 0.01\,  v_w {1\over g_*}\,
        {M_1^4\over T^5\Delta_w}
\label{eq:ans2}
\ee
where $\eta^{\rm raw}_{\sss L}$ is the ratio of $n_L$
to the entropy density s. In the right hand side 
$\Delta_w$ is the wall width and $g_*$ is the 
effective thermodynamic degrees of freedom at the epoch with
temperature $T$.  Using $M_{\sss 1} =f_1 \Delta_{\sss T}$, 
with $\Delta_{\sss T}$ is the temperature
dependent VEV acquired by the $\Delta_{\sss R}$ in the phase
of interest, and $\Delta_w^{-1} = \sqrt{\lambda_{eff}}
\Delta_{\sss T}$ in equation (\ref{eq:ans2}) we get
\be
\eta^{\rm raw}_{\sss B-L} \cong 10^{-4} v_w 
\left(\frac{\Delta_{\sss T}}{T} \right)^5 f_1^4 
\sqrt{\lambda_{eff}}.
\ee
Here we have used $g_*=110$. Therefore, depending on the various
dimensionless couplings, the raw asymmetry can take a range a values
of $O(10^{-4}~-~10^{-10})$. In particular, the cases $f_1<1$ or $f_1\ll1$
implicit in our numerical calculation are compatible with above formula. 
\section{Summary and Conclusion}
In this paper we assume non-thermal production of 
raw lepton asymmetry during the $B-L$ gauge 
symmetry breaking phase transitions. If this asymmetry
passes without much dilution to be the currently observed
baryon asymmetry consistent with WMAP and Big Bang nucleosynthesis,
then the effective neutrino mass parameter
$\tilde{m}_1$ must be less than $10^{-3}eV$. Solution of the
relevant Boltzmann equations shows that 
for $\tilde{m}_1=10^{-4}eV$ the  mass of 
lightest right handed neutrino $N_1$ has to be smaller than 
$10^{12}GeV$ and can be  as low as $10$ TeV. 
In a more restrictive scenario where the 
neutrino Dirac mass matrix is identified with that of 
the charged leptons it is necessary that $M_1>10^8 GeV$ 
in order to satisfy $\tilde{m}_1 < 10^{-3}eV$. Therefore in 
the more restricted scenario all values $M_1$, $10^8 GeV < 
M_1 < 10^{12}GeV$ can successfully create the required 
asymmetry. If the Dirac mass scale of neutrinos is less restricted,
much lower values of $M_1$ are allowed. In particular, 
a right handed neutrino as low as 10 TeV is admissible.

If the raw asymmetry is large, the numerical solutions
show a small window for $M_1$ to get the final asymmetry 
of the required order. The allowed range gets smaller 
as the raw asymmetry gets larger. This is true 
for all allowed values of the neutrino mass parameter 
$\tilde{m}_1$. 

In summary, if the $B-L$ gauge symmetry is gauged, we start
with a clean slate for $B-L$ number and an asymmetry in it can be generated
by a non-perturbative mechanism at the scale where it breaks.
The presence of heavy right handed neutrinos still permits
sufficient asymmetry to be left over in the form of baryons
for a large range of values of the $B-L$ breaking scales.
While other mechanisms of leptogenesis become unnatural
below $10^8$ GeV this mechanism even tolerates TeV scale.
A specific mechanism of this kind is possible in the context of
Left-Right symmetric model, presumably embedded in the
larger unifying group $SO(10)$.  Upon incorporation of supersymmetry,
the qualitative picture remains unaltered. In the simplest
possibility, the scale of SUSY breaking is higher than the
$B-L$ breaking scale, 
in which case the present considerations will carry through
without any change. The gravitino bound of $10^9$ GeV for
reheating temperature after inflation is easily accommodated.

\section{Acknowledgment} NS wishes to thank M. Plumacher for 
helpful discussion and N. Mohapatra for her help in plotting 
fig. \ref{fig:regions}.


\end{document}